\documentclass[prl,twocolumn,floats,superscriptaddress]{revtex4-1}
\usepackage{amsmath}
\usepackage{amssymb}
\usepackage{graphicx}
\usepackage{color}

\newcommand{\red}[1]{\textcolor{red}{#1}}

\newcommand{\xx}{\ensuremath{\boldsymbol{x}}}

\begin{document}

\title{A reduction methodology for fluctuation driven population dynamics}
\date{\today}

\author{Denis S.\ Goldobin}
    \affiliation{ Institute of Continuous Media Mechanics, Ural Branch of RAS, Acad. Korolev street 1, 614013 Perm, Russia}
    \affiliation{ Department of Theoretical Physics, Perm State University, Bukirev street 15,
614990 Perm, Russia}
\author{Matteo di Volo }
     \affiliation{Laboratoire de Physique Th\'eorique et Mod\'elisation, Universit\'e de Cergy-Pontoise,CNRS, UMR 8089, 95302 Cergy-Pontoise cedex, France}
\author{Alessandro Torcini}
                \email[corresponding author: ]{alessandro.torcini@cyu.fr}
    \affiliation{Laboratoire de Physique Th\'eorique et Mod\'elisation, Universit\'e de Cergy-Pontoise,CNRS, UMR 8089, 95302 Cergy-Pontoise cedex, France}
    \affiliation{CNR - Consiglio Nazionale delle Ricerche - Istituto dei Sistemi Complessi, via Madonna del Piano 10, I-50019 Sesto Fiorentino, Italy}

        \date{\today}

\begin{abstract}
Lorentzian distributions have been largely employed in statistical
mechanics to obtain exact results for heterogeneous systems.
Analytic continuation of these results is impossible even
for slightly deformed Lorentzian distributions, due to the divergence of all the moments (cumulants).
We have solved this problem by introducing a {\it pseudo-cumulants'} expansion.
This allows us to develop a reduction methodology for heterogeneous spiking neural networks subject to
extrinsinc and endogenous noise sources, thus generalizing the mean-field formulation introduced in
[E. Montbri\'o {\it et al.}, Phys. Rev. X 5, 021028 (2015)].
\end{abstract}

\maketitle

\paragraph{Introduction}

The Lorentzian ({\it or} Cauchy) distribution (LD) is the second most important stable distribution
for statistical physics (after the Gaussian one)~\cite{zolotarev1986}, which can be expressed in a
simple analytic form, i.e.
\begin{equation}
L(y)=\frac{\pi^{-1}\Delta}{\Delta^2+(y-y_0)^2}
\label{LD}
\end{equation}
where $y_0$ is the peak location and $\Delta$ is the half-width at half-maximum (HWHM).
In particular, for a heterogenous system with random variables distributed accordingly to a LD
it is possible to estimate exactly the average observables
via the residue theorem ~\cite{yakubovich1969}.

This approach has found large applications in physics, ranging from quantum optics,
where it was firstly employed to treat in exactly the presence of heterogeneities
in the framework of laser dynamics~\cite{lamb1964,yakubovich1969}, to condensed matter, where the
Lloyd model~\cite{lloyd1969} assumed a LD for the potential disorder to obtain exact results for the Anderson
localization in a three-dimensional atomic lattices~\cite{anderson1958}.
Furthermore, thanks to a Lorentzian formulation exact results can be obtained for various problems related
to collective dynamics of heterogeneous oscillator populations \cite{rabinovich1989,crawford1994}.
Moreover, LDs emerge naturally for the phases of self-sustained oscillators driven by common noise
\cite{goldobin2005,goldobin2019}.

More recently, the Ott--Antonsen (OA) Ansatz~\cite{ott2008,ott2009} yielded closed
mean-field (MF)  equations for the dynamics of the synchronization order parameter for globally coupled phase oscillators on the basis of a wrapped LD of their phases. The nature of these phase elements can vary from the phase reduction of biological and chemical oscillators~\cite{winfree1967, kuramoto2003} through superconducting Josephson junctions~\cite{watanabe1994,marvel2009} to directional elements like active rotators~\cite{dolmatova2017,klinshov2019} or magnetic moments~\cite{tyulkina2020}.

A very important recent achievement has been the application of the OA Ansatz to heterogeneous globally coupled networks of spiking neurons, namely of quadratic integrate-and-fire (QIF) neurons \cite{luke2013, laing2014}. In particular, this formulation has allowed to derive a closed low-dimensional set of macroscopic equations describing exactly the evolution of the population firing rate and of the mean membrane potential~\cite{montbrio2015}. In the very last years the Montbri\'o--Paz\'o--Roxin (MPR) model \cite{montbrio2015}
is emerging as a representative of a new generation of neural mass models able to successfuly capture relevant aspects of
neural dynamics \cite{devalle2017,byrne2017, dumont2017, devalle2018, schmidt2018network,coombes2019, pietras2019, dumont2019, ceni2020cross,segneri2020theta,taher2020, montbrio2020}.

However, the OA Ansatz (as well as the MPR model) is not able to describe the presence of random fluctuations,
which are naturally present in real systems due to noise sources of different nature.
In brain circuits  the neurons are sparsely connected and {\it in vivo} the presence of noise is unavoidable
\cite{gerstner2014}. These fundamental aspects of neural dynamics have been successfully
included in high dimensional MF formulations of spiking networks based on Fokker-Planck or self consistent approaches
\cite{brunel1999,brunel2000,schwalger2017}. A first attempt to derive a low dimensional MF model for
sparse neural networks has been reported in \cite{volo2018,bi2020}, however the authors mimicked the effects of the random connections only in terms of quenched disorder by neglecting endogenous fluctuations in the synaptic inputs. Fluctuations which have been demonstrated to be essential for the emergence of collective behaviours in recurrent networks \cite{brunel1999,brunel2000}.

In this Letter we introduce a general reduction methodology for dealing with deviations from the LD on the real line,
based on the characteristic function and on its  expansion in {\it pseudo-cumulants}.
This approach avoids the divergences related to the expansion in conventional moments or cumulants.
The implementation and benefits of this formulation are demonstrated for populations of QIF neurons
in presence of extrinsic or endogenous noise sources, where the conditions for a LD of the membrane potentials~\cite{montbrio2015} are violated as in~\cite{volo2018,ratas2019}. In particular, we will derive a hierarchy of low-dimensional MF models, generalizing the
MPR model, for globally coupled networks with extrinsic noise and for sparse random networks with a peculiar focus
on noise driven collective oscillations (COs).

\paragraph{Heterogeneous populations of quadratic integrate-and-fire neurons.}
Let us consider a globally coupled recurrent network of $N$ heterogeneous QIF neurons,
in this case the evolution of the membrane potential $V_j$ of the $j$-th neuron is given by
\begin{equation}
\dot{V_j}=V_j^2+I_j\,, \qquad I_j=I_0+\eta_j+ J_j s(t) + \sigma_j \xi_j(t),
\label{eq101}
\end{equation}
where $I_0$ is the external DC current, $\eta_j$ the neural excitability,
$J_j s(t)$ the recurrent input due to the activity $s(t)$ of the neurons
in the network and mediated by the synaptic coupling of strenght $J_j$.
Furthermore, each neuron is subject to an additive Gaussian noise of amplitude
$\sigma_j = \sigma(\eta_j,J_j)$, where $\langle\xi_j(t)\xi_l(t^\prime)\rangle=2\delta_{jl}\delta(t-t^\prime)$ and $\langle\xi_j\rangle=0$.
The $j$-th neuron emits a spike whenever the membrane potential $V_j$ reaches $+ \infty$
and it is immediately resetted at $-\infty$ \cite{ermentrout1986parabolic}. For istantaneous
synapses, in the limit $N \to \infty$ the activity of the network $s(t)$ will coincide with the
population firing rate $r(t)$ \cite{montbrio2015}. Furthermore, we assume that the parameters
$\eta_j$ ($J_j$) are distributed accordingly to a LD $g(\eta)$ ($h(J)$) with
median $\eta_0$ ($J_0$) and half-width half-maximum $\Delta_\eta$ ($\Delta_J$).

In the thermodynamic limit, the population dynamics can be characterized in terms of the
probability density function (PDF) $w(V,t|\boldsymbol{x})$ with $\boldsymbol{x}=(\eta,J)$,
which obeys the following Fokker--Planck equation (FPE):
\begin{equation}
\frac{\partial w(V,t|\boldsymbol{x})}{\partial t}+\frac{\partial}{\partial V}\Big[(V^2+I_{\boldsymbol{x}})
w(V,t|\boldsymbol{x})\Big]
 =\sigma^2_{\xx}\frac{\partial^2 w(V,t|\boldsymbol{x})}{\partial V^2},
\label{eq102}
\end{equation}
where $I_{\boldsymbol{x}} \equiv I_0+\eta + J r(t)$.
In \cite{montbrio2015}, the authors made the Ansatz that for any initial
PDF $w(V,0|\xx )$ the solution of Eq.~\eqref{eq102} in absence of noise
converges to a LD $w(V,t|\boldsymbol{x})=a_{\xx}/[\pi(a_{\xx}^2+(V-v_{\xx})^2)]$, where $v_{\boldsymbol{x}}$ and
\[
r_{\boldsymbol{x}}(t)=\lim_{V\to\infty}V^2 w(V,t|\boldsymbol{x})=\frac{a_{\boldsymbol{x}}}{\pi} \quad,
\]
represent the mean membrane potential and the firing rate
for the $\xx$-subpopulation. This Lorentzian Ansatz has been shown to correspond
to the OA Ansatz for phase oscillators \cite{montbrio2015} and joined with the
assumption that the parameters $\eta$ and $J$ are indipendent and Lorentzian
distributed lead to the derivation of exact low dimensional
macroscopic evolution equations for the spiking network \eqref{eq101}
in absence of noise.

\paragraph{Characteristic function and pseudo-cumulants.}

Let us now show how we can extend to noisy systems the approach derived
in \cite{montbrio2015}. To this extent we should introduce the characteristic function for $V_{\xx}$,
i.e. the Fourier transform of its PDF, namely
\[
\mathcal{F}_{\xx}(k)=\langle{e^{ikV_{\xx}}}\rangle =\mathrm{P.V.}\int\nolimits_{-\infty}^{+\infty}e^{ikV_{\xx}}w(V_{\xx},t|\xx)\mathrm{d}V_{\xx}
\]
in this framework the FPE \eqref{eq102} can be rewritten as
\begin{equation}
\partial_t\mathcal{F}_{\xx}=ik[I_{\xx}\mathcal{F}_{\xx} -\partial_k^2\mathcal{F}_{\xx}]
-\sigma^2_{\xx} k^2\mathcal{F}_{\xx}\, ;
\label{eq103}
\end{equation}
for more details on the derivation see \cite{supp}.
Under the assumption that $\mathcal{F}_{\xx}(k,t)$ is an analytic function of the parameters $\xx$
one can estimate the average chracteristic function for the population $F({\xx},t)=
\int\mathrm{d}\eta \int\mathrm{d}J  \mathcal{F}_{\xx}(\xx,t)g(\eta) h(J) $ and the corresponding FPE via the residue theorem, with the caution that different contours have to be chosen for positive (upper half-plane) and negative $k$ (lower half-plane). Hence, the FPE is given by
\begin{equation}
\partial_t F=ik\left[H_0 F-\partial_k^2F\right]-|k| D_0 F -S_0^2k^2F\,;
\label{eq104}
\end{equation}
where $H_0 = I_0+\eta_0+J_0r$, $D_0 = \Delta_\eta + \Delta_J r$ and
$S_0^2=\sigma^2(\eta_0  +i\Delta_\eta k/|k|, J_0  +i\Delta_J k/|k|)={\cal N}_R+i{\cal N}_I$.
For the logarithm of the characteristic function, $F(k)=e^{\Phi(k)}$, one obtains the following evolution equation
\begin{equation}
\partial_t\Phi=ik[H_0-\partial_k^2\Phi-(\partial_k\Phi)^2]-|k| D_0 -S_0^2k^2.
\label{eq105}
\end{equation}

In this context the Lorentzian Ansatz amounts to set $\Phi_L=ikv-a|k|$ \cite{NL},
by substituting $\Phi_L$ in \eqref{eq105} for $S_0=0$ one gets
\begin{equation}
\dot{v}=H_0+a^2-v^2,
\quad
\dot{a}=2av+D_0\,,
\label{eq:MPR}
\end{equation}
which coincides with the two dimensional MF model found in ~\cite{montbrio2015}
with $r=a/\pi$.

In order to consider deviations from the LD, we
analyse the following general polynomial form for $\Phi$
\begin{equation}
\Phi=-a|k|+ ikv- \sum_{n=2}^\infty \frac{q_n |k|^n+i p_n |k|^{n-1}k}{n}
\qquad .
\label{phi}
\end{equation}
The terms entering in  the above expression are dictated by
the symmetry of the characteristic function ${\cal F}_{\xx}(k)$ for real-valued $V_{\xx}$, which is invariant for a change of sign of $k$ joined to the complex conjugation.
For this characteristic function neither moments, nor cumulats can be determined~\cite{lukacs1970}.
Therefore, we will introduce the notion of {\it pseudo-cumulants}, defined as follows
\begin{equation}
W_1\equiv a-iv\,,\quad W_n\equiv q_n+ip_n\,.
\label{eq107}
\end{equation}
By inserting the expansion \eqref{phi} in the Eq. \eqref{eq105} one gets
the evolution equations for the pseudo-cumulants, namely:
\begin{align}
&\dot{W}_m=(D_0-iH_0)\delta_{1m}+2 ({\cal N}_R + i {\cal N}_I) \delta_{2m}
\nonumber\\
&\qquad{} +im\Big(-mW_{m+1}+\sum\nolimits_{n=1}^{m}W_nW_{m+1-n}\Big) \quad.
\label{eq110}
\end{align}
It can be shown \cite{supp}  that
the modulus of the pseudo-cumulats scales as $|W_m| \propto |S_0|^{2(m-1)}$ with the noise amplitude, therefore it is justified to consider an expansion limited
to the first two pseudo-cumulants. In this case, one obtains the following MF equations
\begin{subequations}
\label{mfnew}
\begin{eqnarray}
\dot{r}&=&\Delta_\eta/\pi+ \Delta_J r+ 2rv+p_2/\pi,
\label{eq113}
\\
\dot{v}&=&I_0+\eta_0+J_0r-\pi^2r^2+v^2+q_2,
\label{eq114}
\\
\dot{q}_2&=&2{\cal N}_R +4(p_3+q_2v-\pi p_2r),
\label{eq115}
\\
\dot{p}_2&=&  2{\cal N}_I + 4(-q_3+\pi q_2r+p_2v).
\label{eq116}
\end{eqnarray}
\end{subequations}

As we will show in the following the above four dimensional set of equations (with the simple closure $q_3=p_3=0$)
is able to reproduce quite noticeably the macroscopic dynamics of globally coupled QIF populations in presence of
additive noise, as well as of deterministic sparse QIF networks.
Therefore the MF model \eqref{mfnew} represents an extention of the MPR model to system subject to either extrinsic or endogenous noise sources.

It can be demonstrated \cite{supp} that the definitions of the firing rate $r=\lim_{V\to\infty}V^2w(V,t)$ and of the mean membrane potential $v=\mathrm{P.V.}\int_{-\infty}^{+\infty}Vw(V,t)\,\mathrm{d}V$
in terms of the PDF $w(V,t)$ obtained in \cite{montbrio2015} for the LD
are not modified even if the PDF includes the correction terms $\{q_n,p_n\}$ .

\paragraph{Globally coupled network with extrinsic noise.}
In order to show the quality of the MF formulation \eqref{mfnew} let us consider
a globally coupled network of QIF neurons each subject to an independent
additive Gaussian noise term of amplitude $\sigma$ (i.e. ${\cal N}_R = \sigma^2$, ${\cal N}_I =0$).
In this framework, we show that the model \eqref{mfnew} reproduces the
macroscopic dynamics of the network in different dynamical regimes relevant for
neural dynamics. Let us first consider the asynchronous dynamics, this amounts to
a fixed point solution $({\bar r},{\bar v}, {\bar q}_2,{\bar p}_2)$ for \eqref{mfnew}.
In this case we can give a clear physical interpretation of the
stationary corrections ${\bar q}_2$ and ${\bar p}_2$. They can be interpreted as
a measure of an  additional source of heterogeneity in the system induced by the noise, indeed the stationary solution of \eqref{mfnew} coincides with those of the MPR model \eqref{eq:MPR} with excitabilities
distributed accordingly to a Lorentzian PDF of median $\eta_0 + {\bar q}_2$  and HWHM $\Delta_\eta + {\bar p}_2$.

\begin{figure}
\begin{center}
\includegraphics[width=1 \linewidth]{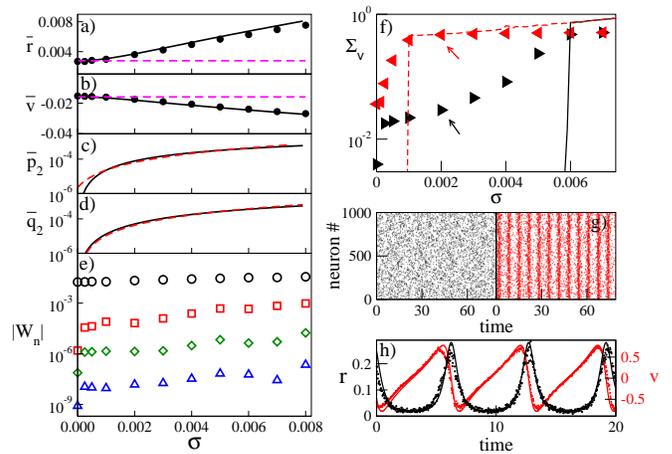}
\end{center}
\caption{{\bf (a-e) Asynchronous Dynamics} Stationary values ${\bar r}$ (a), ${\bar v}$ (b), ${\bar p}_2$ (c), ${\bar q}_2$ (d)  and
$|W_n|$ (e) versus noise amplitude $\sigma$. (a-b) Symbols refer to network simulations with $N=16000$, solid line to the MF model \eqref{mfnew},  dashed (magenta) lines
are the values of ${\bar r}$ and ${\bar v}$ for the MPR model. In (c-d) the dashed red lines refer to a quadratic fit
to the data. In (e) the symbols refer from top to bottom to
$|W_1|$, $|W_2|$, $|W_3|$ and $|W_4|$. Other parameters : $I_0=0.0001$,$J_0=-0.1$, $\Delta_J=0.1$.
{\bf (f-h) Emergence of COs} (f) Standard deviation $\sum_v$ of $v$ obtained for quasi-adiabatic variation of $\sigma$. Lines (symbols) refer to MF (network) simulations: solid black (dashed red) lines and right (left) triangles are obtained by increasing (decreasing) $\sigma$. (g) Raster plots for a network of $N=32000$ neurons, only 1000 are displayed. The black and red dots refer to the two coexisting states denoted by arrows of the same color in (f) for $\sigma=0.002$. (h) $r$ ($v$) versus time for $\sigma=0.002$: dots refer to network simulations with $N=32000$ and lines to MF results. Other parameters: $I_0=0.38$, $J_0=-6.3$ and $\Delta_J=0.01$. In all cases
$\eta_0= \Delta_\eta=0$.}
\label{fig1}
\end{figure}

As shown in Fig. \ref{fig1} (a-b), in the asynchronous regime the MF model \eqref{mfnew} reproduces quite well the population firing rate and the mean membrane potential obtained by the network simulations, furthermore as
reported in Fig. \ref{fig1} (c-d) the corrections $q_2$ and $p_2$ scales as $\propto \sigma^2$ as expected.
The truncation to the second order of the expansion \eqref{eq110} which leads to \eqref{mfnew}
is largely justified in the whole range of noise amplitude here considered. Indeed as displayed
in Fig. \ref{fig1} (e) $|W_1| \sim {\cal O} (10^{-2})$ and $|W_2| \sim {\cal O} (10^{-4})$,
while the moduli of the other pseudo-cumulants are definitely smaller.

For a different set of parameters, characterized by stronger recursive couplings and higher external DC currents,
we can observe the emergence of COs. This bifurcation from asynchronous to coherent behaviours can be characterized
in term of the standard deviation $\sum_v$ of the mean membrane potential: in the thermodynamic limit $\sum_v$ is
zero (finite) in the asynchronous state (COs). As shown in Fig. \ref{fig1} (f)
the MF model reveals a hysteretic transition from the asynchronous state to COs characterized by
a sub-critical Hopf bifurcation occurring at $\sigma_{\rm HB} \simeq 0.0055$.
The coexistence of asynchronous dynamics and COs is observable in a finite range delimted on one side by $\sigma_{\rm HB}$ and on the other by a saddle-node bifurcation of limit cycles at $\sigma_{\rm SN} \simeq 0.00095$.
This scenario is confirmed by the network simulations with $N=64000$ (triangles in panel (f)), however due to the finite size effects $\sum_v$ is not expected to vanish in the asynchronous state. The coexistence of
different regimes is well exemplified by the raster plots shown in panel (g).
The comparison of the simulations and MF data reported in Fig. \ref{fig1} (h)
show that the model \eqref{mfnew} is able to accurately reproduce
the time evolution of $v$ and $r$  also during COs.

\paragraph{Sparse networks exhibiting endogenous fluctuations.} Let us now consider a sparse random network characterized by
a LD of the in-degrees $k_j$ with median $K$ and HWHM $\Delta_k = \Delta_0 K$,
this scaling is assumed in analogy with an exponential distribution. By following \cite{brunel1999}, we can
assume at a MF level that each neuron $j$ receives
$k_j$ Poissonian spike trains characterized by a rate $r$, this amounts to have
an average synaptic input $\frac{J_0 k_j}{K} r(t)$ plus Gaussian fluctuations of
variance $\sigma^2_j = \frac{J_0^2 k_j r(t)}{2K^2}$.
Therefore, as shown in  \cite{volo2018} the quenched disorder in the
connectivity can be rephrased in terms of a random
synaptic coupling. Namely, we can assume that the neurons are fully coupled,
but with random distributed synaptic couplings
$J_j = \frac{J_0 k_j}{K}$  with median $J_0$ and HWHM $\Delta_J = J_0 \Delta_0$.
Furthermore, each neuron $j$ will be subject to a noise of variance $\sigma_j^2 = \frac{J_0 J_j}{2K} r(t)$
and this amounts to have ${\cal N}_R = \frac{J_0^2 r}{2K}$ and ${\cal N}_I = - \frac{J_0^2 \Delta_0 r }{2K}$.

\begin{figure}
\begin{center}
\includegraphics[width=1 \linewidth]{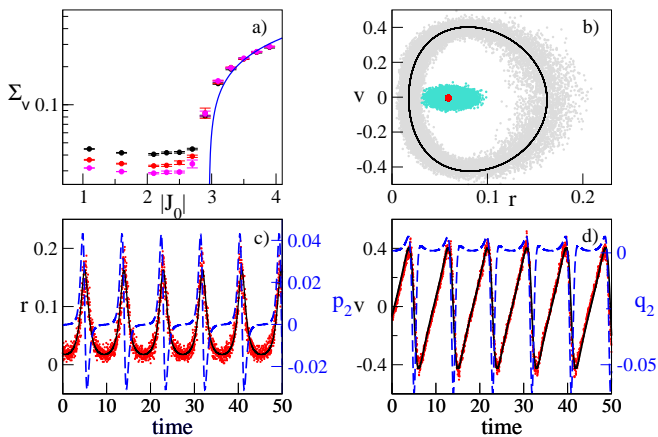}
\end{center}
\caption{{\bf Sparse Network} (a) Standard deviation $\sum_v$ of $v$ versus the synaptic coupling $|J_0|$. Blue solid line (colored symbols) refers to MF (network) simulations. The different colors black, red and magenta correspond to different network sizes $N=10000$, 15000 and 20000, respectively. The error bars are estimated as the standard deviations over 8 different realisations of the random network. (b) Attractors in the $r$-$v$ plane
for $J_0=-2.5$ (asynchronous) and $J_0=-3.7$ (collective oscillations). For the latter value of the coupling
the MF evolution of $r$ ($p_2$) and $v$ ($q_2$) is displayed in (c) and (d) as black solid (blue dashed) lines, respectively.
In (c-d) the symbols refer to the network simulations with $N=40000$. In all panels the parameters are $K=4000$,$\Delta_J=0.01$, $I_0=0.19.$ and $\eta_0= \Delta_\eta=0$. The standard deviations $\sum_v$ are estimated over a time window of $T=500$ after discarding a transient of the same duration.
}
\label{fig2}
\end{figure}

By considering the MF model \eqref{mfnew} for this random network and
by increasing the synaptic coupling, we observe a supercritical Hopf transition
from asynchronous to oscillatory dynamics at $J_0 \simeq 2.956$.
The standard deviation $\sum_v$ of $v$ is reported
in Fig. \ref{fig2} (a) for the MF and for network simulations of different
sizes for $K=4000$. The network simulations confirm the existence of a
transition to collective behaviour for $J_0 \simeq 2.9$.
Furthermore, the asynchronous and coherent attractors in the $r$-$v$ plane
 observed for the MF and the network simulations are in good agreement, despite
the finite size effects (as shown in Fig. \ref{fig2} (b)).
Finally, in presence of collective oscillations the time
evolution for $r$ and $v$ obtained by the network simulations
are well reproduced by the MF dynamics (see Fig. \ref{fig2} (c) and (d)).

It should be remarked that the inclusion of the quenched disorder
due to the heterogeneous in-degrees in the MPR model is not sufficient to lead
to the emergence of COs, as shown in \cite{volo2018}.
It is therefore fundamental to take in account corrections to the
Lorentzian Ansazt due to endogenous fluctuations.
Indeed, as shown in Fig. \ref{fig2} (c) and (d)
the evolution of $r$ and $v$ is clearly guided by that of
the corrective terms $p_2$ and $q_2$ displaying
regular oscillations.

\paragraph{Conclusions.} A fundamental aspect that renders the LD $L(y)$ \eqref{LD}
difficult to employ in a perturbative approach is that all moments and cumulants diverge, which holds
true also for any distribution with $y^{-2}$-tails.
However, to cure this aspect one can deal with the characteristic function of $y$ and introduce an expansion in {\it pseudo-cumulants}, whose form is suggested from the LD structure.
As we have shown, this expansion can be fruitfully applied to build low dimensional MF reductions
for QIF spiking neural networks going beyond the MPR model~\cite{montbrio2015}, since
our approach can also encompass different types of noise sources.
In particular, the MPR model is recovered by limiting the expansion to the
first pseudo-cumulant. Moreover, the stability of the MPR manifold
can be rigorously analyzed within our framework by considering higher order pseudo-cumulants.

Our approach allows one to derive in full generality a hierarchy of
low-dimensional  neural mass models able to reproduce, with the desidered accuracy, firing rate and
mean membrane potential evolutions for heterogeneous sparse populations of QIF neurons.
Furthermore, our formulation applies also to populations of identical neurons in the limit
of vanishing noise~\cite{devalle2018,laing2018}, where the macroscopic dynamics is attracted to
a manifold that is not necessary the OA (or MPR) one \cite{goldobin2020}.

One of the main important aspects of the MPR formulation, as well as of our
reduction methodology, is the ability of these MF models to capture transient
synchronization properties and oscillatory dynamics present in the spiking
networks \cite{devalle2017,schmidt2018network,coombes2019, taher2020}, but that are lost
in usual rate models as the Wilson-Cowan one \cite{wilson1972excitatory}.
Low dimensional rate models able to capture the synchronization dynamics
of spiking networks have been recently introduced \cite{schaffer2013, pietras2020},
but they are usually limited to homogenous populations.
MF formulations for heterogeneous networks subject to extrinsic noise sources
have been examined in the context of the  {\it circular cumulants} expansion
\cite{tyulkina2018,goldobin2019b,ratas2019,pietras2020}.  However,
as noticed in \cite{goldobin2019b},
this expansion has the drawback that any finite truncation leads to
a divergence of the population firing rate.
Our formulation in terms of pseudo-cumulants does not suffer of these strong limitations and
as shown in \cite{supp} even the definition of the macroscopic observables
is not modified by considering higher order terms in the expansion.

Potentially, the introduced framework can be fruitfully applied  to one-dimensional models of Anderson localization, where the localization exponent obeys a stochastic equation similar to Eq. (\ref{eq101}) \cite{lifshitz1988} and also to achieve generalizations of the 3D Lloyd model~\cite{lloyd1969}, of the theory of heterogeneous broadening of the laser emission lines~\cite{yakubovich1969}, and of some other problems in condensed matter and collective phenomena theory involving
heterogenous ensembles.

\begin{acknowledgments}
We acknowledge stimulating discussions with Lyudmila Klimenko, Gianluigi Mongillo, Arkady Pikovsky, and Antonio Politi.
The development of the basic theory of pseudo-cumulants was supported by the Russian Science Foundation (Grant No. \ 19-42-04120).
A.T.\ and M.V. received financial support by the Excellence Initiative I-Site Paris Seine (Grant No.\ ANR-16-IDEX-008), by the Labex MME-DII (Grant No.\ ANR-11-LBX-0023-01), and by the ANR Project ERMUNDY (Grant No.\ ANR-18-CE37-0014), all part of the French program Investissements d’Avenir.
\end{acknowledgments}

\bibliographystyle{apsrev4-1}

\begin{widetext}
\section*{Supplemental Material on \\
``A reduction methodology for fluctuation driven population dynamics''}
\subsection*{by Denis S.\ Goldobin, Matteo di Volo, and Alessandro Torcini
}

%
%
%
%
%

\section{Characteristic function and pseudo-cumulants}

Here we report in full details the derivation of the model (11), already outlined in the Letter,
in terms of the characteristic function and of the associated pseudo-cumulants. In particular,
the characteristic function for $V_{\xx}$ is defined as
\[
\mathcal{F}_{\xx}(k)=\langle{e^{ikV_{\xx}}}\rangle =\mathrm{P.V.}\int\nolimits_{-\infty}^{+\infty}e^{ikV}w(V,t|{\xx})\,\mathrm{d}V \;,
\]
which for a Lorentzian distribution becomes\,:
\[
\mathrm{P.V.}\int\nolimits_{-\infty}^{+\infty}e^{ikV}\frac{a_{\xx}}{\pi[a_{\xx}^2+(V-v_{\xx})^2]}\mathrm{d}V
=e^{ikv_{\xx}-a_{\xx}|k|} \;.
\]

In order to derive the FPE in the Fourier space, let us proceed with a more rigourous
definition of the characteristic function, namely
$$
\mathcal{F}_{\xx}\equiv\lim_{\varepsilon\to+0}\langle{e^{ikV_{\xx}-\varepsilon|V_{\xx}|}}\rangle \;.$$
Therefore by virtue of the FPE (Eq.~(3) in the Letter) the time derivative of the characteristic function takes the form
\begin{align}
\partial_t \mathcal{F}_{\xx} =\lim\limits_{\varepsilon\to+0}\mathrm{P.V.}\int\limits_{-\infty}^{+\infty}e^{ikV_{\xx}-\varepsilon|V_{\xx}|} \frac{\partial w_{\xx}}{\partial t} \mathrm{d}V_{\xx} =
-\lim\limits_{\varepsilon\to+0}\mathrm{P.V.}\int\limits_{-\infty}^{+\infty}e^{ikV_{\xx}-\varepsilon|V_{\xx}|}
\frac{\partial }{\partial V_{\xx}}\left((I_{\xx}+V_{\xx}^2)w_{\xx}-\sigma_{\xx}^2\frac{\partial}{\partial V_{\xx}} w_{\xx} \right)\mathrm{d}V_{\xx}
\nonumber\\
=-\lim\limits_{\varepsilon\to+0}\lim\limits_{B\to+\infty}\int\limits_{-B}^{B}e^{ikV_{\xx}-\varepsilon|V_{\xx}|}
\frac{\partial }{\partial V_{\xx}}\left((I_{\xx}+V_{\xx}^2)w_{\xx}-\sigma_{\xx}^2\frac{\partial}{\partial V_{\xx}} w_{\xx} \right)\mathrm{d}V_{\xx}
 \;.
\nonumber
\end{align}
Performing a partial integration, we obtain
\begin{equation}
\partial_t \mathcal{F}_{\xx} =-\lim\limits_{\varepsilon\to+0}\lim\limits_{B\to+\infty}
\left(
\left.e^{ikV_{\xx}-\varepsilon|V_{\xx}|}q_{\xx}(V_{\xx})\right|_{-B}^{B}
-\int\limits_{-B}^{B}
\frac{\partial e^{ikV_{\xx}-\varepsilon|V_{\xx}|}}{\partial V_{\xx}}q_{\xx}(V_{\xx})\mathrm{d}V_{\xx}
\right) \;,
\label{eq:dFx}
\end{equation}
where the probability flux for the ${\xx}$-subpopulation is defined as
\[
q_{\xx}=(I_{\xx}+V_{\xx}^2)w_{\xx}-\sigma_{\xx}^2\frac{\partial w_{\xx}}{\partial V_{\xx}}
\;.
\]

As the membrane potential, once it reaches the threshold $+B$, is reset to $-B$ this sets
a boundary condition on the flux, namely $q_{\xx}(B)=q_{\xx}(-B)$ for $B\to+\infty$; therefore,
$$
e^{ikB-\varepsilon B}q_{\xx}(B)-e^{-ikB-\varepsilon B}q_{\xx}(-B)
=2ie^{-\varepsilon B}\sin{kB}\,q_{\xx}(B)
\stackrel{B\to+\infty}{\longrightarrow}0
$$
and the first term in Eq.~(\ref{eq:dFx}) will vanish, thus the time derivative of the characteristic function is simply given by
\begin{equation}
\partial_t \mathcal{F}_{\xx} =\lim\limits_{\varepsilon\to+0}\lim\limits_{B\to+\infty}
\int\limits_{-B}^{B}
ike^{ikV_{\xx}-\varepsilon|V_{\xx}|}\left((I_{\xx}+V_{\xx}^2)w_{\xx} -\sigma_{\xx}^2\frac{\partial w_{\xx}}{\partial V_{\xx}}\right)\mathrm{d}V_{\xx}
 \;.
\nonumber
\end{equation}

Hence, after performing one more partial integration for the remaining $V_{\xx}$-derivative term, we obtain
\begin{eqnarray}
\partial_t\mathcal{F}_{\xx} &=&\lim\limits_{\varepsilon\to+0}\mathrm{P.V.}\int\limits_{-\infty}^{+\infty} e^{ikV_{\xx}-\varepsilon|V_{\xx}|}
\left[ ik\left(I_{\xx}+V_{\xx}^2\right)w_{\xx} -\sigma^2_{\xx} k^2w_{\xx}\right]\mathrm{d}V_{\xx}
\nonumber
 \\
&=& ik\left(I_{\xx}\mathcal{F}_{\xx} +\lim\limits_{\varepsilon\to+0}\mathrm{P.V.}\int\limits_{-\infty}^{+\infty} e^{ikV_{\xx}-\varepsilon|V_{\xx}|} V_{\xx}^2 w_{\xx} \mathrm{d}V_{\xx} \right)
-\sigma^2_{\xx}  k^2\mathcal{F}_{\xx}
\end{eqnarray}
and finally
\begin{equation}
\partial_t\mathcal{F}_{\xx}=ik[I_{\xx}\mathcal{F}_{\xx} -\partial_k^2\mathcal{F}_{\xx}]-\sigma_{\xx}^2k^2\mathcal{F}_{\xx}\;,
\label{eqS103}
\end{equation}
which is Eq.~(4) in the Letter.

Under the assumption that $\mathcal{F}_{\xx}(k,t)$ is an analytic function of the parameters $\xx$
one can calculate the average characteristic function for the population $F(k,t)=
\int\mathrm{d}\eta \int\mathrm{d}J  \mathcal{F}_{\xx}(k,t)g(\eta) h(J) $ and the corresponding FPE via the residue theorem, with the caution that different contours have to be chosen for positive (upper half-planes of complex $\eta$ and $J$) and negative $k$ (lower half-planes). Hence, the FPE is given by
\begin{equation}
\partial_t F=ik\left[H_0 F-\partial_k^2F\right]-|k| D_0 F -S_0^2k^2F\;,
\label{eqS104}
\end{equation}
where $H_0 = I_0+\eta_0+J_0r$, $D_0 = \Delta_\eta + \Delta_J r$ and
$S_0^2=\sigma^2(\eta_0  +i\Delta_\eta k/|k|, J_0  +i\Delta_J k/|k|)=\mathcal{N}_R+i\mathcal{N}_I$.

For the logarithm of the characteristic function, $F(k)=e^{\Phi(k)}$, one obtains the following evolution equation
\begin{equation}
\partial_t\Phi=ik[H_0-\partial_k^2\Phi-(\partial_k\Phi)^2]-|k| D_0 -S_0^2k^2\;.
\label{eqS105}
\end{equation}

In this context the Lorentzian Ansatz amounts to set $\Phi_L=ikv-a|k|$ \cite{NL},
by substituting $\Phi_L$ in \eqref{eq105} for $S_0=0$ one gets
\begin{eqnarray}
\dot{v}&=&H_0+a^2-v^2\;,
\nonumber \\
\dot{a}&=&2av+D_0\;,
\label{eqS:MPR}
\end{eqnarray}
which coincides with the two dimensional mean-field model found in~\cite{montbrio2015}
with $r=a/\pi$.

In order to consider deviations from the Lorentzian distribution, we
analyse the following general polynomial form for $\Phi$\,:
\begin{equation}
\Phi=-a|k|+ ikv- \sum_{n=2}^\infty \frac{q_n |k|^n+i p_n |k|^{n-1}k}{n}
\quad .
\label{Sphi}
\end{equation}
The terms entering in  the above expression are dictated by
the symmetry of the characteristic function ${\cal F}_{\xx}(k)$ for real-valued $V_{\xx}$, which is invariant for a change of sign of $k$ joined to the complex conjugation.
For this characteristic function neither moments, nor cumulats can be determined~\cite{lukacs1970}.

Hence, we can choose the notation in the form which would be most optimal for our consideration. Specifically, we introduce
$\Psi=k \partial_k\Phi$,
\begin{equation}
\Psi=-(a\mathrm{sign}(k)-iv)k-(q_2+ip_2\mathrm{sign}(k))k^2 -(q_3 \mathrm{sign}(k)+ i p_3)k^3-\dots\;.
\label{eqS106}
\end{equation}
Please notice that
\begin{align}
\Psi(-k)=\Psi^\ast(k)\quad\left[\mbox{as well as }\Phi(-k)=\Phi^\ast(k)\right].
\label{eqS108}
\end{align}
In this context Eq.~(\ref{eqS105}) becomes
\begin{equation}
\partial_t\Psi=ik H_0-|k|D_0-ik\partial_k\left(k\partial_k\frac{\Psi}{k}+\frac{\Psi^2}{k}\right) -2S_0^2k^2\;.
\label{eqS109}
\end{equation}

It is now convenient to introduce the {\it pseudo-cumulants}, defined as follows:
\begin{equation}
W_1\equiv a-iv\;,\qquad W_n\equiv q_n+ip_n\;.
\label{eqS107}
\end{equation}

From Eq.~(\ref{eqS109}) we can thus obtain the evolution equation for the pseudo-cumulants $W_m$, namely
\begin{equation}
\dot{W}_m=(D_0-iH_0)\delta_{1m}+2(\mathcal{N}_R+i\mathcal{N}_I)\delta_{2m} +im\Big(-mW_{m+1}+\sum\nolimits_{n=1}^{m}W_nW_{m+1-n}\Big)\;,
\label{eqS110}
\end{equation}
where for simplicity we have assumed $k>0$ and employed the property~(\ref{eqS108}). Moreover,
we have omitted the $k\delta(k)$ contribution, since it vanishes.

The evolution of the first two pseudo-cumulant reads as:
\begin{align}
&\dot{W}_1=D_0-iH_0-iW_2+iW_1^2\;,
\label{eqS111}
\\
&\dot{W}_2=2(\mathcal{N}_R+i\mathcal{N}_I) +4i(-W_3+W_2W_1)\;.
\label{eqS112}
\end{align}

Or equivalently
\begin{subequations}
\label{Smfnew}
\begin{eqnarray}
\dot{r}&=&\Delta_\eta/\pi+ \Delta_J r+ 2rv+p_2/\pi\;,
\label{eqS113}
\\
\dot{v}&=&I_0+\eta_0+J_0r-\pi^2r^2+v^2+q_2\;,
\label{eqS114}
\\
\dot{q}_2&=&2{\cal N}_R +4(p_3+q_2v-\pi p_2r)\;,
\label{eqS115}
\\
\dot{p}_2&=&  2{\cal N}_I + 4(-q_3+\pi q_2r+p_2v)\;,
\label{eqS116}
\end{eqnarray}
\end{subequations}
which is Eq.~(11) in the Letter.

\section{Firing rate and mean membrane potential for perturbed Lorentzian distributions}

In the following we will demonstrate that the definitions of the firing rate $r$ and of the mean membrane
potential $v$ in terms of the PDF $w(V,t)$, namely:
$$
r=\lim_{V\to\infty}V^2w(V,t) \qquad {\rm and} \qquad
v=\mathrm{P.V.}\int_{-\infty}^{+\infty}Vw(V,t)\,\mathrm{d}V \; ,
$$
obtained in \cite{montbrio2015} for a Lorentzian distribution,
are not modified even by including in the PDF the correction terms $\{q_n,p_n\}$\;.

The probability density for the membrane potentials $w(V,t)$ is related to the characteristic function
$F(k)$ via the follwoing anti-Fourier transform
$$w(V,t)=(2\pi)^{-1}\int\nolimits_{-\infty}^{+\infty} F(k)\,e^{-ikV}\mathrm{d}k$$
with $F(k)=e^{\Phi(k)}$. By considering the deviations of $\Phi(k)$ from the Lorentzian distribution
up to the seond order in $k$, we have
\begin{eqnarray}
2 \pi \enskip w(V,T) &=& \int\nolimits_{-\infty}^{+\infty} e^{ikv-a|k|-q_2\frac{k^2}{2}-ip_2\frac{k|k|}{2}}e^{-ikV}\mathrm{d}k \nonumber \\
&\approx & \int\nolimits_{-\infty}^{+\infty} e^{-iky-a|k|}\left( 1-q_2\frac{k^2}{2}-ip_2\frac{k|k|}{2}\right)\mathrm{d}k
\nonumber\\
&=&\int\nolimits_{-\infty}^{+\infty}\bigg(
 1+\frac{q_2}{2}\left[(1-\theta)\frac{\partial^2}{\partial y^2}-\theta\frac{\partial^2}{\partial a^2}\right]
 -\frac{p_2}{2}\frac{\partial^2}{\partial y \partial a}\bigg)e^{-iky-a|k|}\mathrm{d}k\;,
\nonumber
\end{eqnarray}
where $y=V-v$ and $\theta$ is an arbitrary parameter. Thus one can
rewrite
\begin{equation}
w(V,t)\approx\bigg(1
 +\frac{q_2}{2}\left[(1-\theta)\frac{\partial^2}{\partial y^2}
 -\theta\frac{\partial^2}{\partial a^2}\right] {}-\frac{p_2}{2}\frac{\partial^2}{\partial y \partial a}\bigg)\frac{a}{\pi(a^2+y^2)}\;.
\label{eqB01}
\end{equation}

From the expression above, it is evident that $q_2$ and $p_2$, as well as the higher-order corrections, do not modify the firing rate definition reported in \cite{montbrio2015} for the
Lorentzian distribution, indeed
\[
r=\lim_{V\to\infty}V^2w(V,t)=\frac{a}{\pi}\;.
\]

Let us now estimate the mean membrane potential by employing the PDF~(\ref{eqB01}),
where we set the arbitrary parameter $\theta$  to zero without loss of generality,
namely
\[
w(V,t)=\left(1
 +\frac{q_2}{2}\frac{\partial^2}{\partial V^2}
 -\frac{p_2}{2}\frac{\partial^2}{\partial V \partial a}+\dots\right)w_0(V,t)\;,
\]
where $w_0(V,t)=\pi^{-1}a/[a^2+(V-v)^2]$\,. The mean membrane potential is given by
\begin{eqnarray}
\langle{V}\rangle&=&\mathrm{P.V.}\int_{-\infty}^{+\infty}Vw(V,t)\,\mathrm{d}V
=\mathrm{P.V.}\int_{-\infty}^{+\infty}\left(Vw_0
 -\frac{q_2}{2}\frac{\partial w_0}{\partial V}
 +\frac{p_2}{2}\frac{\partial w_0}{\partial a}+\dots\right)\mathrm{d}V
\nonumber\\
&=&v-\frac{q_2}{2}\int_{-\infty}^{+\infty}\frac{\partial w_0}{\partial V}\mathrm{d}V
 +\frac{p_2}{2}\frac{\partial}{\partial a}\int_{-\infty}^{+\infty}w_0\mathrm{d}V+\dots
\nonumber\\
&=&v-\frac{q_2}{2}w_0|_{-\infty}^{+\infty}
 +\frac{p_2}{2}\frac{\partial\, 1}{\partial a}+\dots=v\;.
\label{eqB02}
\end{eqnarray}

All the higher-order corrections enetering in $w(V,t)$, denoted by $(\dots)$ in \eqref{eqB02}, have the form of higher-order derivatives of $w_0$ with respect to $V$ and $a$; therefore they yield a zero contribution to the estimation of $\langle{V}\rangle$. Thus, Eq.~(\ref{eqB02}) is correct not only to the 2nd order, but also for higher orders of accuracy. We can see that the interpretation of the macroscopic variables $a=\pi r$ and $v=\langle{V}\rangle$ in terms of the firing rate and of the mean membrane potential entering in
 Eq.~(\ref{eqS110}) or Eqs.~(\ref{eqS113})--(\ref{eqS116}) remains exact even away from the Lorentzian distribution.

\section{Smallness hierarchy of the pseudo-cumulants}

Eq.~(\ref{eqS110}) for $m>1$ can be recast in the following form
\begin{equation}
\dot{W}_{m>1}=2m(v+i\pi r)W_m+
2({\cal N}_R + i {\cal N}_I) \delta_{2m} +im\Big(-mW_{m+1}+\sum\nolimits_{n=2}^{m-1}W_nW_{m+1-n}\Big) \;,
\label{eqH01}
\end{equation}
where $W_m$ is present only in the first term of the right-hand side of the latter equation.

Let us now understand the average evolution of $W_M$. In particular, by dividing Eq.~(\ref{eq113}) by $r$ and averaging over time, one finds that
\begin{equation}
\overline{v}=-\frac{\Delta_\eta +\overline{p_2}}{2\pi\overline{r}}-\frac{\Delta_J}{2}\;,
\label{eqH02}
\end{equation}
where $\overline{\cdot}$ denotes the average in time and where we have employed the fact that
the time-average of the time-derivative of a bounded process is zero, i.e.\ $\overline{\frac{\mathrm{d}}{\mathrm{d}t}\ln{r}}=0$. Since $r(t)$ can be only positive, $\overline{v}$ will be strictly negative for a heterogeneous population ($\Delta_\eta \ne 0$ and/or $\Delta_J \ne 0$) in the case of nonlarge deviations from the Lorentzian distribution, i.e., when $p_2$ is sufficiently small.
In particular, for asynchronous states $v=\overline{v}$, hence, Eq.~(\ref{eqH02}) yields a relaxation dynamics for $W_m$ under forcing by $W_{m+1}$ and $W_1,\dots,W_{m-1}$; by continuity, this dissipative dynamics holds also for oscillatory regimes which are not far from the stationary states.

Let us explicitly consider the dynamics of the equations~(\ref{eqH01}) for $m=2,3,4$, namely:
\begin{align}
\dot{W}_2&=4(v+i\pi r)W_2-i4W_3+2 ({\cal N}_R + i {\cal N}_I)\;,
\label{eqH03}
\\
\dot{W}_3&=6(v+i\pi r)W_3+i3W_2^2-i9W_4\;,
\label{eqH04}
\\
\dot{W}_4&=8(v+i\pi r)W_4+i8W_2W_3-i16W_5\;,
\label{eqH05}
\\
&\dots\;.
\nonumber
\end{align}
In absence of noise terms ${\cal N}_R={\cal N}_I=0$, we consider a small deviation from the Lorentzian distribution
such that $|W_n|< C \varepsilon^{n-1}$, where $C$ is some positive constant and $\varepsilon\ll1$ is a smallness parameter. In this case, from Eq.~(\ref{eqH03}) one observes that $W_2$ tends to $\sim W_3$, while
from Eq.~(\ref{eqH04}), $W_3\to\sim W_2^2$. Therefore, $W_2\to\sim W_2^2$, which means that $W_2(t\to+\infty)\to0$.
Further, from Eq.~(\ref{eqH05}), $W_4\to\sim W_2W_3\sim W_2^3\to0$. Thus, in absence of noise, the systems tends to a state $W_1 \ne 0$, $W_{m>1}=0$ (at least from a small but finite vicinity of this state). This tell us that the Lorentzian distribution is an attractive solution in this case.

In presence of noise, by assuming that $|{\cal N}_R + i {\cal N}_I| \sim \sigma^2$, a similar analysis of Eqs.~(\ref{eqH03})--(\ref{eqH05}) yields $|W_2| \to\sim\sigma^2$, $|W_3| \to \sim |W_2^2| \sim\sigma^4$, \dots,

$$|W_m|\to\sim\sigma^{2(m-1)} \;.$$

The above scaling is well confirmed by the data reported in Fig.~\ref{figS1}.
Therefore, a two-element truncation~(\ref{eqS111})--(\ref{eqS112}) of the infinite equation chain~(\ref{eqS110})
is well well justified as a first significant correction to the Lorentzian distribution dynamics.

\begin{figure}
\begin{center}
\includegraphics[width=0.5 \linewidth]{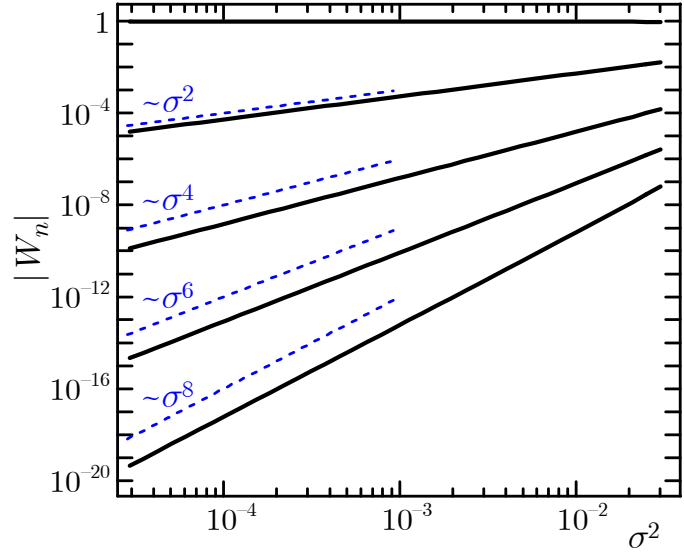}
\end{center}
\caption{Modulus of the pseudo-cumulants $|W_n|$ versus the noise variance $\sigma^2$ for
$n$ ranging from 1 to 5 from top to bottom.  The pseudo-cumulants are estimated
by integrating Eq.~(10) in the main text with extended precision (30 digits)
and by limiting the sum to the first 100 elements.
Other parameters: $I_0=0.1$, ${\eta}_0=-1$, $J_0=1$, $\Delta_{\eta} = 0.1$ and $\Delta_J=0.1$.}
\label{figS1}
\end{figure}

\end{widetext}

\end{document}